\documentclass[preprint,showpacs,amsmath,amssymb]{revtex4}

\usepackage{amsmath}
\usepackage{graphicx}
\usepackage{epsfig}

\newcommand{\be}{\begin{equation}}
\newcommand{\ee}{\end{equation}}
\newcommand{\bma}{\begin{displaymath}}
\newcommand{\ema}{\end{displaymath}}

\begin{document}
\title{Electron-hole duality and vortex rings in quantum dots}

\author{M. Manninen$^1$, S.M. Reimann$^2$, M. Koskinen$^1$, Y. Yu$^2$ and M. Toreblad$^2$}

\affiliation{\sl $^1$NanoScience Center, Department of Physics,
FIN-40014 University of Jyv\"askyl\"a, Finland}

\affiliation{\sl $^2$Mathematical Physics, Lund Institute of Technology,
SE-22100 Lund, Sweden}
 
\date{2004-10-11}

\begin{abstract} 
In a quantum-mechanical system, 
particle-hole duality implies that 
instead of studying {\it particles}, 
we can get equivalent 
information by studying the missing particles, the so-called {\it holes}. 
Using this duality picture for rotating fermion condensates the 
vortices appear as holes in the Fermi see. 
Here we predict that 
the formation of vortices in quantum dots at high magnetic fields
causes oscillations in the energy spectrum
which can be experimentally observed using accurate tunnelling 
spectroscopy. 
We use the duality picture to show 
that these oscillations are caused by the localisation of 
vortices in rings. 
\end{abstract}
\pacs{71.10.-w, 73.21.La, 05.30.Fk, 03.75.Lm}

\maketitle

Vortex formation in quantum liquids has been much discussed
in connection with finite rotating condensates of bosonic 
atoms~\cite{madison2000,abo,butts,ben,bertsch,schweikhard2004}. 
More recently, even fermionic 
atom gases~\cite{rodriquez2001,bulgac2003,sbeouelji2001} entered the stage.
Quite generally, and independent of the underlying statistics, 
it appears that vortex formation is a universal property 
of rotating 
quantal systems with repulsive interaction among the 
particles~\cite{toreblad2004}. 
Analogously, theoretical studies have demonstrated that vortices are formed in 
semiconductor quantum dots at strong external magnetic
fields~\cite{saarikoski2004}, or equivalently at
high angular momenta.

Particle-hole duality is frequently used for understanding quantum
mechanical many-particle systems: 
Sometimes the study of the missing particles, i.e. holes,
gives a much clearer view of the observed phenomena.
For example, when describing the conductance of a semiconductor 
the use of holes greatly simplifies a complicated 
physical picture~\cite{ashcroft1976}. 
Related concepts of duality have been used in nuclear 
physics, where states within 
some given angular momentum shell can be described either by 
particles or, equivalently, holes in the same shell~\cite{bohr1969}.
A similar description was used already in the 1930's for the atomic 
shell model~\cite{shortley1932}. Recently, particle-hole
duality has also been studied in connection 
with quantum Hall liquids~\cite{girvin1996,shahar1996,burgess2001}.

In this letter we show, starting from exact many-particle theory,
and using general arguments that (i) vortices in quantum dots are holes 
in the polarised Fermi see of electrons, 
(ii) vortices form a fractional quantum Hall system,
(iii) vortices localise in a ring, and
(iv) the energy spectrum as a function of magnetic field shows 
oscillations which can be measured and reveal the number of 
vortices in the dot.

We first note that in a fairly strong magnetic field
the electrons confined by the circular quantum dot form a polarised
electron gas. The role of the magnetic field is only to set the electron cloud to 
rotation~\cite{reimann1999,manninen2001}. For the present 
purpose it is thus sufficient to study the development of 
the many-body state of the electrons with increasing total 
angular momentum $L$. 

Figure 1 shows the energy spectrum for 20 electrons in a circular
two-dimensional quantum dot as a function of $L$. 
The spectrum is the result of exact diagonalisation of
the Hamiltonian describing the system of interacting electrons~\cite{remark1}.
We observe that the lowest-energy state as a function of angular momentum
shows periodic oscillations with {\it successive periods of 2, 3, and 4}. 
As we will explain below, these oscillations are 
{\it signatures of 2, 3 and 4 vortices being localised in a ring.} 

Unfortunately, the accuracy of the resonant tunneling measurements 
available today ~\cite{oosterkamp1999,reimann2002} 
is not sufficient to uncover these oscillations of the grund state 
energies with increasing angular momentum and magnetic field.
However, the very recent developments in sample 
preparation and accuracy of tunneling spectroscopy~\cite{bjork}
in fact should make it possible to observe these salient features. 

To our surprise, we found that the reason behind the periodic 
oscillations in the energy spectrum is deeply connected 
with the above mentioned particle-hole duality.

Let us consider a two-dimensional harmonic confinement 
with $N$ interacting electrons. 
In a highly rotating state, for which the orbital 
angular momentum reaches $L=N(N-1)/2$, the so-called maximum
density droplet~\cite{macdonald1993} is formed. This state is the finite-size
manifestation of the integer quantum Hall liquid. When it occurs, the 
electron system is fully polarised~\cite{remark2}.
Consequently, we can restrict our study to the many-particle physics
of {\it spinless} electrons in high rotational states.
We use the configuration interaction (CI) technique to solve the 
many-particle Schr\"odinger equation. The many-particle 
configuration is a Slater determinant of single particle
states of the two-dimensional harmonic oscillator. 
Beyond the maximum density droplet the configuration space can
be restricted to the lowest Landau band~\cite{manninen2001}.
Any configuration can then be described with a vector
$\vert n_0,n_1,\cdots ,n_{M}\rangle$, where $n_m$ is the occupation
number (0 or 1) for the single particle state with angular 
momentum $m$, and
$M$ is the the highest single particle angular momentum included in
the basis. A state with $N$ electrons will 
have $N$ {\it one}'s and $(M-N)$ {\it zero}'s
in the configuration vector, i.e., it will for example be of the form
$\mid 111110001111\cdots \rangle $.
We call the {\it one}'s particles and the {\it zero}'s holes.
The key issue to be noted here is that 
{\it the many-particle problem can be solved by diagonalizing the 
Hamiltonian matrix either for the particle states or for the hole states},
i.e., replacing each $n_m$ with $1-n_m$ (the above Fock state
will then be $\mid 000001110000\cdots \rangle $ for holes)\cite{remark3}.
This duality of the rotational states has two advantages:
(i) by {\it interpreting the holes as vortices},
it gives a natural way to study the vortex-vortex correlations, i.e. 
the internal structure of the vortex lattice, 
and (ii) it shows that for small number of vortices, the vortex
state can be approximated with the Laughlin wave function for the
fractional quantum Hall state, while the electron state is 
close to the integer quantum Hall state. Although the electrons
remain delocalized, the holes will localise
in rings resulting in characteristic oscillations of the energy spectra 
as a function of angular momentum or, equivalently, magnetic field. 

Figure 2 shows the vortex-vortex pair correlation for 20 electrons
with angular momentum 246. In Fig. 2a the actual calculation for
{\it 20 electrons} was performed using $M=30$ single
particle states.  {\it After} diagonalisation of the Hamiltonian matrix, 
the {\it many-particle} 
state was converted to a {\it few-hole} state, simply by converting 
each $n_m$ to $1-n_m$. The vortex-vortex pair correlation 
clearly shows four localised vortices
in the quantum dot, with three maxima each corresponding to a single vortex,
and one ``missing'' vortex due to the reference point (which resembles the
``exchange hole'' in the particle-particle correlation function). 
The ring surrounding the localised vortices is due to the six 
holes at high single particle angular momentum states 
(for $M=30$ and $N=20$ the total number of 
holes is $M-N=10$). Figure 2b shows the result of the analogous full many-body
calculation for the corresponding situation with  
{\it four holes}. Clearly, the vortex structure is identical
to that of Fig. 2a, except that now trivially, the hole density outside 
the electron cloud has vanished.  
Figure 2 demonstrates clearly that the result of the many-particle
state is similar whether it is done with electrons or with holes. 
A direct comparison of the configuration amplitudes of many-particle 
states of these two calculations indeed 
show a very good agreement~\cite{remark3}. We emphasise that
when diagonalizing the many-body Hamiltonian in the restricted Hilbert space 
of the lowest Landau level, the computational work is not
easier whether using holes or electrons. However, the duality 
description makes it possible to determine the vortex texture via the
vortex-vortex correlation.

Let us now view the many-particle state for holes in terms of   
a fractional quantum Hall state. As an example,  we use the same state
for which we showed the correlation functions in Fig 2b. 
For four holes, the angular momentum of this state now equals 30. 
This appears as a small number compared to the angular momentum
of the conjugate state for 20 electrons ($L=246$). 
In effect, however, the angular momentum of hole states relative to the 
compact maximum density droplet is larger than 
that of the electron states due to the much smaller number of ``particles''.
In fact, the hole state can be well approximated 
by the Laughlin wave function
$\Psi=\prod_{i,j} (z_i-z_j)^q\exp(-\sum_k \vert z_k\vert^2)$ 
(where the electron coordinates are described by 
complex numbers $z_i$)~\cite{laughlin1983}.
Since the angular momentum of such a state is $qN(N-1)/2$,
our four-vortex state with angular momentum 30 corresponds 
to $q=5$. More generally, we can argue that
a small number of holes (vortices) in a rotating system
of a large number of electrons naturally leads to
a Laughlin state with large $q$.

The Laughlin state localises the particles in a lattice.
This localisation is the better
the larger the exponent $q$~(see Refs.~\cite{laughlin1983,manninen2001}).
For less than six vortices in a quantum dot, these vortices are found to 
localise in a ring, similarly to the well-known 
localisation of {\it particles} in a so-called Wigner molecule~\cite{reimann2002}.
{\it It is this localisation of vortices that results in the characteristic 
features of the rotational spectrum.} Rigid rotation of
the ring-like 'molecule' formed by the localised vortices becomes possible.
For polarised fermions, 
as in this case, the rigid rotation of the ring 
of $n$ vortices is allowed only 
at every $n$th angular momentum~\cite{manninen2001,viefers2004,koskinen2001}.
These angular momenta represent the minima in
the energy spectrum, seen in Fig. 1. At intermediate
angular momenta, the rigid rotation has to be accompanied 
with other excitations, like vibrational modes, resulting
in higher energies. Consequently, the period of the successive
minima gives directly the number of the vortices localised in a ring.
The single-vortex state moves towards the center with increasing 
angular momentum, but does not reach the origin before the two-vortex state 
becomes the ground state.

When the vortex number increases to six, the localised
vortices can have two nearly degenerate textures:
A ring of six vortices or a ring of five vortices with
one vortex at the centre. As an example we study six 
vortices in a 30 electron quantum dot. Figure 3 shows 
the vortex-vortex correlation functions for the two
lowest energy states at angular momentum 555, which 
for the conjugate state for holes corresponds 
to the Laughlin state with $q=5$.
The lowest energy state shows a ring of six vortices while
the nearly degenerate first excited state shows a
vortex structure with one vortex at the centre.
Classically, particles interacting with repulsive long-range
interaction (Coulombic or logarithmic) will have these two
stable configurations~\cite{reimann2002}. Quantum mechanical
calculations for spinless fermions also find these 
internal symmetries, as shown in Fig. 3. As a consequence, 
the energy spectrum as a function of the angular momentum
becomes more complicated when the number of vortices 
reaches six, due to the two possible periods of rigid 
rotation~\cite{manninen2001}.

In conclusion, we have shown that the vortex formation
in quantum dots most easily can be understood in terms
of electron-hole duality. The vortex texture is a natural
consequence of the Laughlin-type many-particle wave function 
for vortices. The localisation of vortices in 
a ring appears in the rotational spectrum as 
periodic oscillations of the ground state energy as
a function of external magnetic field. This should also be observable 
experimentally.
Finally, we point out that a related duality holds also
for trapped bosons and explains the vortex lattices in
rotating bose condensates.

{\it Acknowledgements}
This work was supported by the Academy of Finland, the Swedish Foundation 
for Strategic Research and the Swedish Research Council. 


\begin{figure}[h]
\centerline{\epsfxsize=5in\epsfbox{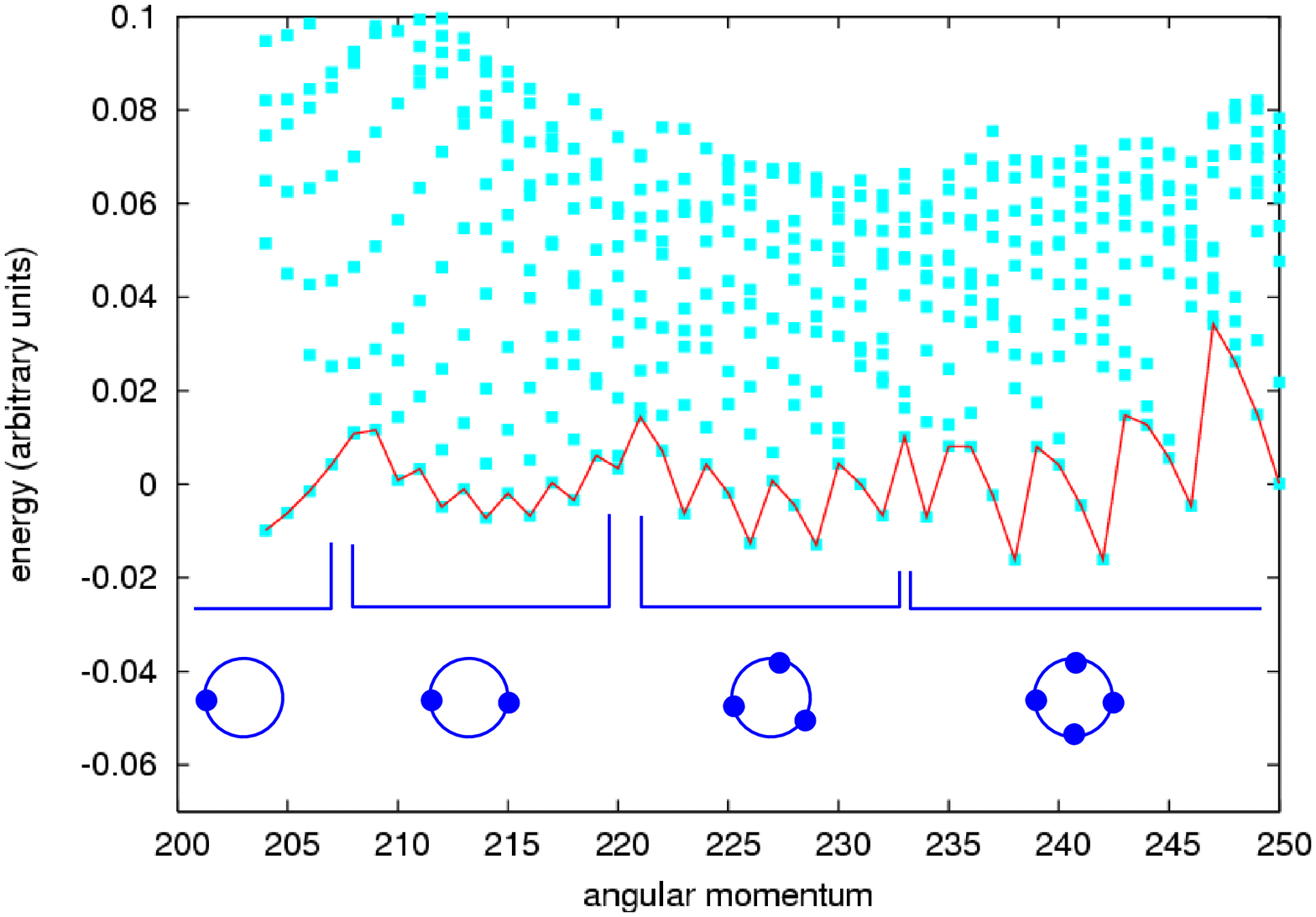}}
\caption{
Energy spectrum of 20 electrons in a quantum dot 
as a function of the angular momentum. The solid
line connects the lowest energy states for each 
angular momentum and shows the characteristic
oscillation caused by the vortex localization.
The vortex textures are shown schematically 
below the regions for 1 to 4 vortices. 
(A smooth function of the angular momentum,
$f(L)=A+BL+CL^2$, was subtracted from
the energies to mimic the effect of the magnetic 
field on the total energy).}
\end{figure}

\begin{figure}[h]
\centerline{\epsfxsize=5in\epsfbox{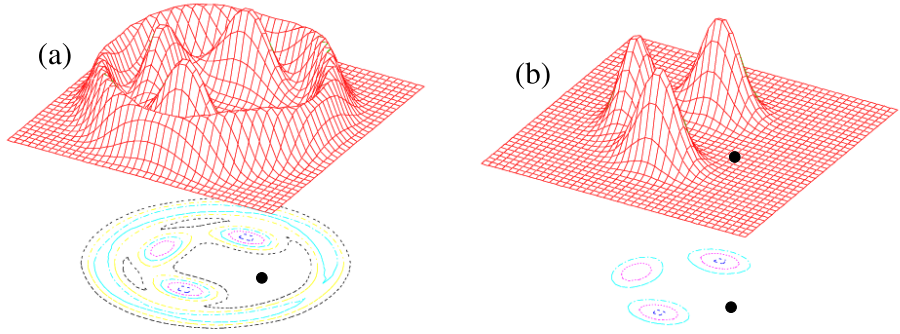}}
\caption{
Vortex-vortex pair correlation function for 
the four-vortex state in a quantum dot with 
20 electrons. The reference point is marked with 
a black dot. (a) shows the result for the calculation
of 20 electrons with angular momentum 246 and 
(b) shows the result for the calculation of the 
conjugate system with 4 holes.}
\end{figure}

\begin{figure}[h]
\centerline{\epsfxsize=5in\epsfbox{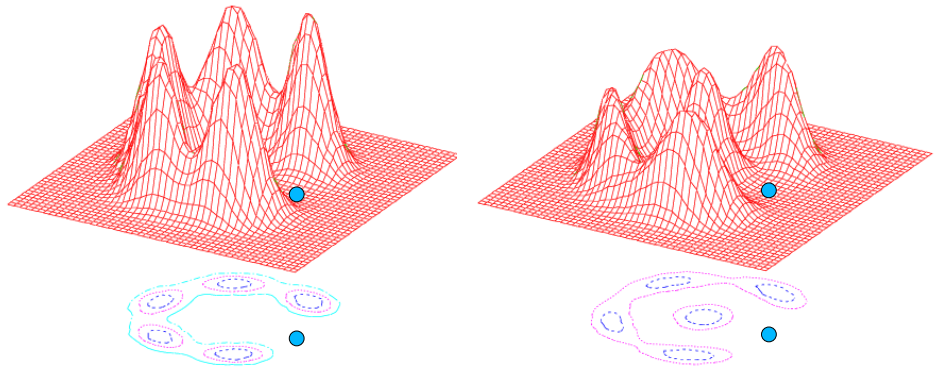}}
\caption{
Vortex-vortex correlations of the two lowest energy
states for 30 electron quantum dot at angular momentum
555. The ground state (a) shows 6 vortices localised in
a ring, while the first excited state (b) shows
a five vortex ring with one vortex in the centre.} 
\end{figure}


\begin{thebibliography}{23}
\bibitem{remark1} The Fock space was constructed of 40 single particle states
of the lowest Landau band and Lanzos technique was used for the 
diagonalization of the Large Hamiltonian matrix.
\bibitem{remark2}
Although, the rotational motion is 
achieved with an external magnetic field, the 
interaction of the field with the electron spin,
i.e. the Zeeman energy, is not needed for the 
polarization~\cite{reimann1999}.
\bibitem{remark3} In our case the particle-hole duality
is only approximative since the diagonal terms of the 
Hamiltonian matrix are different. However, this 
has only small shifts in the energy levels, but no
effects on the qualitative features of the spectrum.
\end{thebibliography}
\end{document}